\documentclass{article}

\usepackage{multirow}
\usepackage{amsmath}
\usepackage{amsfonts}

\newcommand{\pnt}[1]{\boldsymbol{#1}}
\DeclareMathOperator{\dif}{d}

\begin{document}
\title{The impact of phonon dispersion on thermodynamical properties in
computational models of crystalline solids} \author{Roman Belousov
(belousov.roman@gmail.com) \and Mauro Prencipe (mauro.prencipe@unito.it) \\
University of Torino, Department of Earth Sciences,\\
via Valperga Caluso 35, 10125, Turin, Italy,\\
tel. +39-011-6705131, fax +39-011-2365131 } \maketitle \abstract{The existing
techniques of account for the phonon dispersion are computationally costly,
while its impact on a variety of thermodynamic properties appears negligible. We
develop a mathematical formalism, which allows for clear understanding of the
effect. The theoretical elaborations are then confronted with a widely used
phenomenological model of the dispersion relation. The results show that
accuracy of the models, which neglect the phonon dispersion, allows for
calculation of many thermodynamic quantities up to an admissible precision.
Although the context refers mainly to the Helmholtz energy , other properties
are concerned as well with less details. Time-efficient schemes of control over
the accuracy loss due to the neglected phonon dispersion are discussed.}

{\bf keywords}: phonon dispersion, thermodynamic properties, Helmholtz energy
\section{Introduction}
The phonon dispersion represents a challenging problem in computational models
of crystalline solids. As the analytical form of the dispersion relation is so
far not established, the numerical techniques are inevitably come into use. The
computational time demanded by these methods is by few orders longer than the
one needed for calculation of the phonon frequencies solely at $\Gamma$-point.
Despite the impressing power of modern computer resources, this is still quite
costly calculation, e.g. in the quantum-mechanical treatment of problems.

On other hand, in practice it is often met, that some thermodynamic properties
remain unaffected to the precision of interest by the phonon dispersion. That
is, in such case the model restrained to consideration of the crystal unit cell
and its $\Gamma$-point frequencies would lead to a sufficient level of accuracy
and decrease drastically the computational cost. Indeed in many theoretical
studies the phonon dispersion is neglected for these reasons.

Nonetheless the question of reliability and validity of such approximation is
sometimes doubted on fair scientific grounds, as seemingly there was given
neither theoretical justification for this approach, neither clear theoretical
notion on the impact of phonon dispersion on thermodynamic properties. In this
work we present a purely theoretical study of phonon dispersion that lets
conceive the problem on mathematical grounds and provide with altogether missing
criteria to judge on its role in computational models. The elaborated
mathematical treatment further will be applied to the existing phenomenological
model widely used in both experimental and theoretical studies to give the
understanding of why many thermodynamic properties appear insensitive to the
phonon dispersion. Therefrom a theoretical justification for the discussed
approximation follows.

All the reported work is performed with the Helmholtz energy expression as a
sample thermodynamic quantity. However the mathematical statements will be kept
as general as possible, thus the analogous treatment could be freely applied to
other properties with minimal modifications. In Sec.~\ref{sec:discussion} the
concern on other thermodynamic properties and the boundaries of the
approximation will be discussed.

\section{Theoretical background}
\label{sec:background}
The concept of phonon dispersion arises from the cyclic periodic model of atomic
lattice \cite{Maradudin1963,Evarestov2007}. That is, the whole lattice is
represented as a pattern specified by a chosen crystal cell of $p$ atoms, which
is repeated $N$ times by translation vectors. The commonly adopted harmonic
model with Born-von Karman boundary condition leads to the description of nuclei
motion in terms of $3 p$ (3-dimensional case) normal modes and $N$ allowed wave
vectors. Together they form $3 p N$ degrees of freedom. Further 6 normal modes
(3 of translation and 3 of rotation) should be subtracted to yield the number of
genuine vibrations \cite{Cotton1971}. For the model of infinite crystal, i.e.
$N \to \infty$, the number of vibrational modes is $3 p - 3$, as there are no
rotations external to the infinite lattice. The dependence of the vibrational
frequency $\nu_i$ of $i$-th normal mode on the wave vector $\pnt{\kappa}$, which
varies over $N$ values allowed by the boundary condition, is called the phonon
dispersion relation. Its analytical form for a general 3-dimensional case is so
far not established.

All allowed wave vectors can be found within the first Brillouin Zone (BZ),
which is the reciprocal to the chosen cell of direct periodic lattice. The other
vectors are symmetrically equivalent to those of the first BZ due to the
periodicity of the dual lattice. When $N$ tends to infinity, the uniformly
spread wave vectors are treated as a 3-dimensional continuum.

According to the foundations of statistical physics the frequencies of vibratory
motion can be used for evaluation of thermodynamic properties. As stated in the
introduction, we'll consider specifically the expression for Helmholtz energy $F
= E_{st} + F_{vib} + E_{el}$ \cite{Anderson1995}, where $E_{st}$, $F_{vib}$ are
the static lattice energy and the vibrational energy of nuclei, $E_{el}$ is the
term arising due to the presence of free electrons. In common practice these
quantities are obtained per crystal cell (designated by the superscript
\emph{cc}): $F^{cc} = F / N = E_{st} / N + F_{vib} / N + F_{el} / N$. Similar
forms of account for the vibrational motion occur in expressions for other
thermodynamic potentials and properties, which will be briefly touched solely at
the end of manuscript.

Regularly for insulating solids $E_{el}$ is excepted from consideration. The
only term, which concerns the frequencies of normal modes, is $F_{vib}$:
\begin{equation}
F_{vib} = F_{ZP} + F_{Th} = \frac{h}{2} \sum_{i=1}^{3 p N} \nu_i + k T
\sum_{i=1}^{3 p N}
\ln[1 - \exp(-\frac{h \nu_i}{k T})]\text{,}
\end{equation}
where $F_{ZP}$ and $F_{Th}$ are the zero-point and thermal constituents; $h$
and $k$ are the Planck and Boltzmann constants; $T$ is the temperature. We
remark, that the summation runs over all normal modes, including the
translational ones, which have zero frequency. Obviously this does not affect
the zero-point term. We do not amend the definition of the thermal part, which
then stipulates taking the logarithm of 0, because it is not of importance when
the problem will be brought into continuous case.

Taking into account the phonon dispersion relation the summation over
frequencies can also be recast into the sum over the normal modes and the
allowed wave vectors:
\begin{equation}
\label{eq:extended}
F_{vib} = \sum_{i=1}^{3 p}\sum_{j=1}^{N} \{\frac{h}{2} \nu_i(\pnt{\kappa}_j) + k
T \ln[1 - \exp(-\frac{h \nu_i(\pnt{\kappa}_j)}{k T})]\}\text{.}
\end{equation}
The frequencies of normal modes can be calculated in the least tedious manner at
the zero wave vector \cite{Maradudin1963,Evarestov2007}, also called
$\Gamma$-point. Other points of BZ require more complicated schemes, which
lead to procedures of drastically larger workload. The straightforward approach
is to adopt a larger crystal cell. As it will contain more atoms, the increased
number of normal modes corresponds merely to pieces of the same phonon
dispersion branches in the smaller cell. However due to computational cost there
exists a practice to include only the frequencies of normal modes at
$\Gamma$-point in Eq.~\ref{eq:extended}, thus attempting to approximate the
vibrational part of Helmholtz energy expressed per crystal cell:
\begin{equation}
\label{eq:approximation}
F_{vib}^{cc} \approx \sum_{i=1}^{3 p} \{\frac{h}{2} \nu_i(\pnt{\Gamma}) + k
T \ln[1 - \exp(-\frac{h \nu_i(\pnt{\Gamma})}{k T})]\}\text{.}
\end{equation}

The principle goal of this work is to study the theoretical significance of
approximation given in Eq.~\ref{eq:approximation}. Particularly some emphasis
will be given to its consequences in reconstruction of the dependence of
Helmholtz energy on the cell volume ($V^{cc}$) in the analytical form
$F^{cc}(V^{cc})$. It is obtained by means of curve fitting from the underlying
dependence of $E_{st}$ and $\nu_i$ on volume. This approach gained the name of
quasi-harmonic model \cite{Anderson1995}.

\section{Developing mathematical formalism}
The mathematical formalism developed in this section will be grounded on the
existing treatment of the phonon dispersion in theoretical studies. Up to our
knowledge there exist 2 distinct schemes. The first one (supercell approach) is
perhaps the most straightforward. It follows directly from the underlying theory
surveyed in Sec.~\ref{sec:background}. Another very popular model relies on the
concept of phonon density of states. It was apparently inspired by experimental
works where the signal intensities of measured vibrational frequencies lead to
the natural treatment of problem in terms of density distribution.

We note that seemingly no mathematical formalization of the problem was reported
to reveal the connection of these methods. In this section it will be developed
for the case of Helmholtz energy. Thereon the equivalence of both approaches
will be demonstrated.
 
\subsection{Supercell approach}
\label{sec:supercell}
At first we discuss the supercell approach of accounting for the phonon
dispersion, which is quite  straightforward. Here the superscripts $sc$ and $uc$
will refer to the quantities calculated per supercell and unit cell
respectively. The Helmholtz energy ($F$) of a supercell with dimensions $m_1
\times m_2 \times m_3$ according to Eq.~\ref{sec:background} (without loss of
generality in the following the term due to the free electrons is ignored for
convenience as if it be an insulator) is approximately:
\begin{multline}
F^{sc} = m_1 m_2 m_3 F^{uc} \approx
m_1 m_2 m_3 E_{st}^{uc} + \sum_{i=1}^{3 p m_1 m_2 m_3} \{ \frac{h}{2}
\nu_{i}^{sc} \\
+ k T \ln [1 - \exp(-\frac{h \nu_{i}^{sc}}{k T})] \}
= E_{st}^{sc} + \sum_{i=1}^{3 p m_1 m_2 m_3} f_{i}^{sc}\text{,}
\end{multline}
designating $i$-th mode calculated at the $\Gamma$ point of the supercell by
$\nu_i^{sc}$, its vibrational contribution to the Helmholtz energy by $f_i^{sc}$
and the static lattice energy of cell by $E_{st}$, whereas $p$ is the number of
atoms in the unit cell. The approximate equality becomes exact when $m_1 m_2 m_3 = N$
is the total number of the unit cells in the crystal. I.e. the phonon dispersion
becomes completely accounted for. The supercell approach with $m_1 m_2 m_3 < N$
implies a certain approximate treatment of the phonon dispersion, which will be
now explored.

Restarting from the full account of the phonon dispersion for the unit cell,
we'll use the dependence of the energy $f_i^{uc}(\pnt{\kappa}) =
f(\nu_i^{uc}(\pnt{\kappa}))$ on the wave vector $\pnt{\kappa}$ that identifies a
point in the reciprocal lattice. Onwards the superscript $uc$ at $f_i^{uc}$ and
$\nu_i^{uc}$ will be dropped for brevity, because the respective functions of
the supercell will not be involved further. A stack of Brillouin zones belonging
to a supercell with dimensions $m_1 \times m_2 \times m_3$ divides the first
Brillouin zone (BZ) of the unit cell in $m_1 m_2 m_3$ congruent volumetric
elements. These elements contain an equal number of wave vectors allowed by
Born-von Karman boundary condition $\Delta N = N / (m_1 m_2 m_3)$, as all
vectors are uniformly spread according to the translational symmetry of the
reciprocal lattice. The $\Gamma$ point frequencies of the supercell normal modes
correspond to the frequencies of normal modes at $m_1 m_2 m_3$ distinct wave
vectors of the unit cell first BZ.

For instance, when $m = m_1 = m_2 = m_3$, there are 8 meshes with
$\nu_i(\pnt{\kappa})$ known at 8 points $\pnt{\kappa}_j\text{, }j = 1..8$ laying
on the BZ border. Then the number of wave vectors $\Delta N$ per element $\Delta
\pnt{\kappa}$ (the volume of the supercell BZ) is $\frac{\Delta N}{\Delta
\pnt{\kappa}} = \frac{N}{m_1 m_2 m_3 \Delta \pnt{\kappa}} = \frac{N}{8 \Delta
\pnt{\kappa}}$. With each mesh $\Delta \pnt{\kappa}$ labeled by number $j$, one
can associate the function value $f(\pnt{\kappa}_j)$, as there's only one node
which belongs to that mesh and at which the frequency $\nu_i(\pnt{\kappa}_j)$ is
known. Furthermore in the limit of a supercell $m \to \infty$ spanning the whole
infinite crystal we have $\underset{m \to \infty}{\lim} \Delta \pnt{\kappa} =
0$.

Considering $f_i$ as a continuous function of $\pnt{\kappa}$, a summation over
wave vectors of the unit cell first BZ with volume $V_{BZ}^{uc}$ can be
approximated by a Riemann sum on an equidistant grid:
\begin{multline}
\label{eq:Fvib}
F_{vib}^{uc} = F_{vib} / N
= \frac{1}{N} \sum_{i=1}^{3 p}\sum_{j=1}^{N} f_i(\pnt{\kappa}_j)
\approx \frac{1}{N} \sum_{i=1}^{3 p}\sum_{j=1}^{m_1 m_2 m_3} f_i(\pnt{\kappa}_j)
\Delta N \\
= \sum_{i=1}^{3 p}\sum_{j=1}^{m_1 m_2 m_3} f_i(\pnt{\kappa}_j) \frac{\Delta
N}{N \Delta \pnt{\kappa}} \Delta \pnt{\kappa}
\underset{\Delta \pnt{\kappa} \to 0}{\to} \sum_{i=1}^{3 p}
\underset{V_{BZ}^{uc}}{\int} f_i(\pnt{\kappa}) \frac{\dif \ln N}{\dif
\pnt{\kappa}} \dif \pnt{\kappa} \text{.}
\end{multline}

The integral in Eq.~\ref{eq:Fvib} is the exact
expression of $F_{vib}^{uc}$ in the model of infinite crystal. The quantity
written explicitly
\begin{equation}
\frac{\dif \ln N}{\dif \pnt{\kappa}} = \frac{\dif N}{N \dif \pnt{\kappa}}
= \underset{\substack{\Delta \pnt{\kappa} \to 0 \\ m \to \infty}}{\lim}
\frac{m^{-3}}{\Delta \pnt{\kappa}}
= \underset{\substack{\Delta \pnt{\kappa} \to 0 \\ m \to \infty}}{\lim}
\frac{(V_{BZ}^{uc} / V_{BZ}^{sc})^{-1}}{V_{BZ}^{sc}} \equiv (V_{BZ}^{uc})^{-1}
\end{equation}
is important to preserve the energy units of measure. So it leads to a simpler
expression for infinite crystal with the direct lattice unit cell volume $V^{uc}
= (V_{BZ}^{uc})^{-1}$:
\begin{equation}
\label{eq:main}
F_{vib}^{uc} = V^{uc} \sum_{i = 1}^{3 p}\underset{V_{BZ}^{uc}}{\int}
f_i(\pnt{\kappa})\dif\pnt{\kappa}\text{.}
\end{equation}

For the equidistant grid given by a supercell as discussed above, the Riemann
sum of Eq~\ref{eq:Fvib} now can be reduced to:
\begin{multline}
F_{vib}^{uc} \approx \frac{1}{N} \sum_{i=1}^{3 p}\sum_{j=1}^{m_1 m_2 m_3} f_i
(\pnt{\kappa}_j) \frac{N}{m_1 m_2 m_3} \\
= (m_1 m_2 m_3)^{-1} \sum_{i=1}^{3 p}\sum_{j=1}^{m_1 m_2 m_3}
f_i(\pnt{\kappa}_j)\text{.}
\end{multline}
Such estimation of the integral in Eq.~\ref{eq:main} corresponds to a piecewise
linear approximation of the phonon dispersion relation. In the case of $m = m_1 = m_2 = m_3 = 2$, it
degrades to the simplest linear interpolation. For $m = 1$, the formula is
reduced to the unit cell approach with $\Delta \pnt{\kappa} = V_{BZ}^{uc}$ being
the unit cell BZ itself, i.e. this leads to the expression that neglects the
phonon dispersion:
\begin{equation}
F_{vib}^{uc} \approx V^{uc} \sum_{i=1}^{3 p} f_i(\pnt{\Gamma}) V_{BZ}^{uc}
= \sum_{i=1}^{3 p} f_i(\pnt{\Gamma})\text{.}
\end{equation}

The phonon dispersion can be neglected, when the major part of the hyper-volume
(the integration on BZ volume) under the hyper-surface $f_i(\pnt{\kappa})$ is
$f_i(\pnt{\Gamma}) \times V_{BZ}^{uc}$, that corresponds to the integration of
the flat "horizontal" hyper-surface defined by the function value
$f_i(\pnt{\Gamma})$. As practically the frequencies of optic modes do not
change for various wave vectors the order of magnitude, it can be expected.
Therewith the higher order terms become a small correction, i.e. the
approximation implied is a truncation to a certain precision. This criteria
purely relies on the "flatness" of the vibrational energy dependence on wave
vector, particularly, the absence of soft modes is not required. Hence the role
of the phonon dispersion is guided by the range of the vibrational energy
contribution functions $f$ on the BZ. However the acoustic branches routing at 0
Hz frequency of the translational modes at the $\Gamma$ point still may undergo
large variations of frequency. All this will be discussed properly in
Sec.~\ref{sec:phndisp}.

\subsection{Phonon density of states approach}

A different approach frequently met in theoretical studies involves the phonon
density of states (DoS), which is equivalent to the described above under the
following consideration. We'll normalize DoS $G(\nu) = 3 p \frac{\dif N}{\dif
\nu}$ of the whole crystal to obtain the frequency distribution function
$g(\nu)$ \cite{Maradudin1963}:
\begin{equation}
\frac{1}{3 p N} \underset{-\infty}{\overset{+\infty}{\int}} G(\nu) \dif \nu
= \underset{-\infty}{\overset{+\infty}{\int}} g(\nu) \dif \nu = 1 \text{.}
\end{equation}

The frequency distribution function can be expanded as a sum of the
contributions from individual normal modes $g(\nu) = \sum_{i=1}^{3 p}
\rho_i(\nu)$. The mathematical flow can be fulfilled regarding the one-mode
distributions $\rho_i$ with the summation applied afterwards. We'll further
expand the Riemann sum from Eq.~\ref{eq:Fvib} into a triple summation where the
grid nodes are indexed along the coordinate axes of reciprocal lattice from
$\pnt{\kappa}_{1 1 1} = \pnt{\Gamma}$ to $\pnt{\kappa}_{m_1 m_2 m_3}$:
\begin{multline}
\label{eq:Fvib2}
F_{vib}^{uc} = \frac{1}{N} \sum_{j=1}^{m_1 m_2 m_3} f_i(\pnt{\kappa}_j) \Delta N
\\ = \sum_{a=1}^{m_1}\sum_{b=1}^{m_2}\sum_{c=1}^{m_3}
f(\nu_i(\pnt{\kappa}_{abc}))
\frac{\Delta N}{N} = \sum_{abc} f(\nu_{i,abc}) \frac{\Delta N}{N} \text{.}
\end{multline}

In the sum any 2 indices, e.g. $b$ and $c$, select a row of meshes in the grid.
Thus the double summation on these indices has a limit
\begin{equation}
\sum_{bc} = \sum_{b=1}^{m_2}\sum_{c=1}^{m_3}
\underset{m_2\text{, }m_3 \to \infty}{\to}
\underset{\inf \kappa_1}{\overset{\sup \kappa_1}{\int}} \dif \kappa_1 
\underset{\inf \kappa_2}{\overset{\sup \kappa_2}{\int}} \dif \kappa_2
\end{equation}
($\kappa_1$ and $\kappa_2$ are the coordinates in the reciprocal space), that
can be viewed as a projecting operator that maps its operand defined on the
3-dimensional grid of meshes onto the 1-dimensional array labeled by the
remaining third index $a$. Now we define
\begin{equation}
\Delta \nu_{i, abc}
:= \sup_{\pnt{\kappa} \in \Delta \pnt{\kappa}_{abc}} \nu_i(\pnt{\kappa})
- \inf_{\pnt{\kappa} \in \Delta \pnt{\kappa}_{abc}} \nu_i(\pnt{\kappa})
\end{equation}
and continue to expand the sum of Eq.~\ref{eq:Fvib2}:
\begin{multline}
F_{vib}^{uc} = \sum_{abc} f(\nu_{i,abc}) \frac{\Delta N}{N} \\
= \sum_{a=1}^{m_1} \sum_{bc} f(\nu_{i,abc}) \frac{\Delta N}{N \Delta
\nu_{i,abc}} \Delta \nu_{i,abc}
\underset{\Delta \pnt{\kappa}_{abc} \to 0}{\to}
\underset{\inf \nu_i}{\overset{\sup \nu_i}{\int}} \dif \nu
f(\nu) \rho_i(\nu) \text{.}
\end{multline}

At last taking the summation over the contributions of all normal modes we
arrive to the final expression:
\begin{multline}
F_{vib}^{uc} = \sum_{i=1}^{3 p}
\underset{\inf \nu_i}{\overset{\sup \nu_i}{\int}} \dif \nu
f(\nu) \rho_i(\nu) \\
= \underset{\min_i \inf \nu_i}{\overset{\max_i \sup \nu_i}{\int}} \dif \nu
f(\nu) \sum_{i=1}^{3 p} \rho_i(\nu)
= \underset{\min_i \inf \nu_i}{\overset{\max_i \sup \nu_i}{\int}} \dif \nu
f(\nu) g(\nu) \text{.}
\end{multline}

The problem of integration is then delegated to the estimation of the frequency
distribution function $g(\nu)$. In practice it is obtained from the projection
of all the vibrational frequencies together on the single $\Gamma$ point. As
demonstrated, it is an alternative solution of the integration problem as in the
former approach (Sec.~\ref{sec:supercell}), because it is the same limit of the
Riemann sum of Eq.~\ref{eq:Fvib}. The mesh grid set up in both methods is used
for the direct numerical evaluation of the integral at the former one, while in
the latter it's employed for reconstruction of the intermediate function
$g(\nu)$, which is then used as a weight function (normalized multiplicity) for
the integral of $f(\nu)$.

When the linear interpolation is used to describe the phonon dispersion
relation, which corresponds to the supercell approach of $2 \times 2 \times 2$
dimensions, the frequency distribution function is simply set to the uniform
one:
\begin{equation}
\rho_i = u_i(\nu) = \begin{cases}
	1/(\sup \nu_i - \inf \nu_i) \text{, if } \inf \nu_i \leq \nu \leq \sup \nu_i
	\\
	0 \text{, otherwise}
\end{cases}
\text{.}
\end{equation}
Finally, the expression, which neglects the phonon dispersion, corresponds to a
sharp peak of the density at $\nu_i(\pnt{\Gamma})$, which can be described by
the Dirac delta function $\rho_i(\nu) = \delta(\nu - \nu_i(\pnt{\Gamma}))$. Although
it is physically inaccurate as the infinite discontinuities are not expected $0
\leq \rho_i(\nu) \leq g(\nu) < 1 < \delta(0)$, this approximation is still valid
when the dispersion branch is ``flat``:
\begin{equation}
\label{eq:flatdistr}
f(\nu_i(\pnt{\Gamma})) = \int \dif \nu \rho_i(\nu) f(\nu)
= \int \dif \nu \delta(\nu - \nu_i(\pnt{\Gamma})) f(\nu) \text{.}
\end{equation}

\section{Phonon dispersion integrals}
\label{sec:phndisp}
Here we'll inspect in details the integration expressions derived for the
vibrational contribution of each normal mode given by Eq.~\ref{eq:main}, namely:
\begin{equation}
\label{eq:integral}
V^{uc} \underset{V_{BZ}^{uc}}{\int} \dif \pnt{\kappa} f_i(\pnt{\kappa})
= V^{uc} \underset{V_{BZ}^{uc}}{\int} \dif \pnt{\kappa} f(\nu_i(\pnt{\kappa}))
\text{,}
\end{equation}
which will be referred to as the phonon dispersion integral of $i$-th
vibrational normal mode from here onwards, or simply the phonon dispersion
integral with the summation index dropped. Now we'll note that by virtue of the
mean value theorem, as the function $f_i$ can be regarded continuous in the
closed 3-dimensional interval $\bar{\pnt{V}}_{BZ}^{uc}$ and differentiable in
the open interval $\pnt{V}_{BZ}^{uc}$:
\begin{equation}
\exists \pnt{\bar{\kappa}} \in \pnt{V}_{BZ}^{uc}:~V^{uc}
\underset{V_{BZ}^{uc}}{\int} \dif \pnt{\kappa} f_i(\pnt{\kappa})
= (V_{BZ}^{uc})^{-1}\underset{V_{BZ}^{uc}}{\int} \dif \pnt{\kappa}
f_i(\pnt{\kappa}) = f(\nu_i(\bar{\pnt{\kappa}})) \text{.}
\end{equation}
Therefore another perspective to look at the solution of phonon
dispersion integral is the estimation of a special frequency $\bar{\nu_i}
= \nu_i(\bar{\pnt{\kappa}})$. Its value is approximated by $\bar{\nu_i} \approx
\nu_i(\pnt{\Gamma})$, when the dispersion is neglected.

The energy contribution function of \emph{frequency}, which is
monotonically increasing, in its explicit form is:
\begin{equation}
f(\nu) = f_{ZP}(\nu) + f_{Th}(\nu)
= \frac{h}{2} \nu + k T \ln[1 - \exp(- \frac{h \nu}{k T})] \text{.}
\end{equation}

The zero-point energy function $f_{ZP}$ is linear and doesn't require
complicated treatment, while the thermal energy function $f_{Th}$ is slowly
changing at high frequencies, but it drops down to $-\infty$ when $\nu \to 0$
and thus poses a problem of improper integral in the respective part. The latter
concerns particularly the acoustic modes, which branches of dispersion relation
start from 0 Hz. In fact, when the phonon dispersion is neglected, their
contribution is completely discarded.

A convenient preparation is the variable substitution $x = h \nu / k T$ so that
\begin{multline}
f(x) = k T \frac{x}{2} + k T \ln[1 - \exp(-x)]
= k T \left \{ \ln[\exp(x/2)] + ln[1 - \exp(-x)] \right \} \\
k T \ln[\exp(x/2) - \exp(-x/2)] = k T \ln[2 \sinh(x/2)] \text{,}
\end{multline}
therewith accumulating the integrand terms in one. Let's denote the passing band
of $i$-th normal mode by $I_i = [a_i, b_i]$ with $a_i = \inf
\nu_i(\pnt{\kappa})$ and $b_i = \sup \nu_i(\pnt{\kappa})$. The error $\epsilon_f
= |f_i(\bar{\pnt{\kappa}}) - f_i(\pnt{\Gamma})| < f(b_i) - f(a_i)$ tends to zero
when the measure of passing band interval $\mu(I_i) < \delta_\nu$ is
sufficiently small.

Before proceeding further we'll benefit from some practical considerations. The
atomic vibrational frequencies are commonly found to lay within the Tera-Hertz
order and less, rarely going up to the order of 100 THz. Thus we will fix the
upper bound of frequency variations by one Peta-Hertz, which is seemingly quite
an exaggerated value already. The optic modes, as empirically known, give a
non-zero lower bound of the interval ($a_i > 0 \text{ Hz}$), which is greater by
its absolute magnitude than the upper bound of the acoustic modes. Therefore
we'll set the lower bound of a generic optic mode to $10^{-3}$~THz. The ensemble
of optic modes usually cause a concentration of the total frequency distribution
function roughly in vicinity of $\sim 1$~THz.

Along with the discussed empirical observations, to estimate the generic bounds
of accuracy, when the phonon dispersion is neglected, we'll employ a widely used
successful model for the dispersion relation of vibrational modes proposed by
Kieffer \cite{Kieffer1979}.

\subsection{Acoustic modes}\label{sec:ac}

The acoustic modes as was explicated earlier pose a problem of improper
integration. The corresponding value of phonon dispersion integral
(Eq.~\ref{eq:integral}) essentially depends on the rate with which the frequency
approaches zero. At high frequency range the energy contribution function
$f_{ac}$ is usually as flat as of an optic mode. When the phonon
dispersion is neglected and the acoustic branches are discarded from the
calculations, these terms together can be considered as an additive constant:
this is not of importance for energy quantities (which are usually determined up
to an arbitrary constant). The role of this contribution might be observed
solely when a thermodynamic potential is differentiated, if the value of the
phonon dispersion integral is appreciably changing with the respective variable.
Therewith the approach is suitable under the condition:
\begin{equation}
|\frac{\partial f_{ac}(\bar{\pnt{\kappa}}(V^{uc}))}{\partial V}| \leq
\epsilon_{ac} \text{,}
\end{equation}
which puts the limits on the accuracy of approach. According to Kieffer's model,
the frequency distribution function of an acoustic mode takes a general form:
\begin{equation}
\rho_{ac}(\nu) = \frac{(2/\pi)^3 \arcsin(\nu / b_{ac})}{p (b_{ac}^{2} -
\nu^2)^{1 / 2}} \text{,}
\end{equation}
where $\int_{I_{ac}} \dif \nu \rho_{ac}(\nu) = (3 p)^{-1}$, $p$ is the number of
atoms in the unit cell and $b_{ac}$ is the top frequency of the passing band,
which is the only parameter allowing for variation of the modeling expression.
In Table~\ref{tbl:1} the values of the phonon dispersion integral for acoustic
modes obeying Kieffer's model are given for various parameters $b_{ac}$ and
temperatures.

Once the contribution of acoustic branches is considered as a constant, perhaps
the maximum uncertainty imposed on the vibrational part of Helmholtz energy is
$0.37 \text{ Ha} \cdot 3 p^{-1}$ for the variation of upper bound frequency
$\sim 10^{15} \text{ Hz}$, as seen from Table~\ref{tbl:1}. It leads to the
conclusion that the derivative with respect to frequency is remarkably small.
Admitting such large variation in response to stretching/compression of the
volume is clearly an impractically excessive exaggeration, whereas the factor $3
p^{-1}$ might reduce the discarded contribution by few orders of magnitude.
Indeed such accuracy is superfluous in many kinds of calculations. For instance,
the static lattice energy usually has order of KHa and ensures high precision
with respect to the neglected contribution.

\subsection{Optic modes}

Kieffer's model suggests to describe the whole ensemble of optic branches by the
uniform distribution contributing to the total frequency distribution function
(optic continuum \cite{Kieffer1979}), except for few well identified normal
modes. For the latter the Dirac $\delta$-function is proposed to model their
frequency distribution with generally sufficient accuracy. It can be confronted
directly with Eq~\ref{eq:flatdistr} to confirm that their dispersion relation
leads to a flat function $f_i$, as the dispersion relation itself is flat.
Subtracting the number of such optic modes $q$ from the total, there are $3 p -
3 - q$ modes participating in the optic continuum.

Let's denote the interval of optic continuum as a passing band $I_{op}$. Usually
the dispersion branches are restricted from crossing each other at the same
point $\pnt{\kappa}$ due to the symmetry considerations. Therefore they are
commonly arranged in a stack and bound each other from above and below. This
allows for a rough estimation of the passing band of each optic mode in the
continuum $I_{i} \approx [a_i, a_i + \mu(I_{op}) / (3 p - 3 - q)]$. Therewith a
pragmatic approximation of the error due to the negligence of phonon dispersion
in $i$-th mode might be:
\begin{multline}
\epsilon_{i} = |f_i(\bar{\pnt{\kappa}}) - f_i(\pnt{\Gamma})|
\approx |\frac{\dif f(\nu_i(\pnt{\Gamma}))}{\dif \nu_i}
(\nu_i(\bar{\pnt{\kappa}})-\nu_i(\pnt{\Gamma}))|
\\
\approx |\frac{\dif f(\nu_i(\pnt{\Gamma}))}{\dif \nu_i}| \frac{1}{2}\mu(I_{i})
\approx |\frac{\dif f(\nu_i(\pnt{\Gamma}))}{\dif \nu_i}| \frac{\mu(I_{op})}{2 (3
p - 3 - q)} \text{,}
\end{multline}
which follows from the linear approximation of the dispersion relation. From
simple calculus one can find $|\frac{\dif f(\nu)}{\dif \nu}| < 10^{-12}$ on the
domain $[10^{-3}, 10^{3}]$ THz, while the relation $|\frac{\dif
f(\nu)}{f(\nu) \dif \nu}|$ is $\sim 10^{-10}$ and $\sim 10^{-11}$ at $10^{-3}$
and $10^{3}$ THz, respectively. On multiplication by $\frac{\mu(I_{op})}{2 (3
p - 3 - q)}$ this ensures a certain precision for the Helmholtz energy, when the
phonon dispersion is not taken into account. Such precision is often admissible,
if not superior to the sought one, for a variety of problems addressed in modern
computational studies. For example, in the expression of Helmholtz energy the
expected magnitude of errors in individual constituents of the vibrational term
is strongly subdued by the static lattice energy, which usually has order of KHa.

\section{Practical schemes}
The general considerations presented in Sec.~\ref{sec:phndisp} reason the often
discovered fact, that the phonon dispersion produces negligible effects on the
Helmholtz energy. When the available computational resources are insufficient to
take accurate account for the phonon dispersion, it may also justify the
presumption that the modeled solid does not have peculiar features of the
dispersion relations, which would produce significant deviation of the
calculated quantities. Should the problem be still of concern and require a
control, to spare computational time one can address the problem with the linear
model of phonon dispersion, employing the calculations with $2 \times 2 \times
2$ supercell, because it efficiently gives the bounds of phonon dispersion
branches: larger cells will improve the accuracy less and less notably. If the
aim of calculations relies on multiple evaluations for various parameters of the
system, a scheme can be devised to spare the computational cost as follows.

As example we'll consider the reconstruction of Helmholtz energy dependence on
crystal cell volume in analytical form at constant temperature by curve fitting
procedure. This requires evaluation of $F^{cc}(V^{cc})$ for a set of volumes
$\{V^{cc}_i\}$. One can perform the calculation for the unit cell while
neglecting the phonon dispersion to find the array of points $\{(V^{uc},
F^{uc}(V^{uc}))_i\}$. Thereafter the phonon dispersion can be taken into account
for a subset of values $\{V^{uc}_j\} \subset \{V^{uc}_i\}$ using a supercell to
get $\{(V^{uc}, (m_1 m_2 m_3)^{-1} F^{sc}(m_1 m_2 m_3 V^{uc}))_j\}$. Then the
error due to the phonon dispersion is estimated as:
\begin{equation}
\delta F^{uc}(V^{uc}_i) = (m_1 m_2 m_3)^{-1} F^{sc}(m_1 m_2 m_3 V^{uc}_i)
- F^{uc}(V^{uc}_i) \text{.}
\end{equation}
This error can be then interpolated: i) complicated curve fitting procedures are
unlikely to be much of use, given that there are few points $\{V^{uc}_j\}$); ii)
it is important that the points $\{V^{uc}_j\}$ cover the whole studied domain of volume, i.e. $\min_i \{V^{uc}_i\} \in \{V^{uc}_j\}$ and $\max_i
\{V^{uc}_i\} \in \{V^{uc}_j\}$. Besides that the interpolated error $\delta
F(V)$ can provide with estimations at any point on the domain of volume, it
allows for control over the accuracy of inferred thermodynamic quantities, such
as pressure $P = - \partial F / \partial V$, the isothermal bulk modulus $K_0 =
- V \frac{\partial P}{\partial V}$ and others, as will be explicated further.

If the interpolated error is considered as a finite variation of Helmholtz
energy function of volume, simple formulae for the variation of derived
quantities follow:
\begin{align}
\delta P[F(V)] = P[F(V)+\delta F(V)] - P[\delta F(V)]
= - \frac{\partial \delta F(V)}{\partial V} \\
\delta K[F(V)] = V \frac{\partial^2 \delta F(V)}{\partial V ^2}
\end{align}
Alternatively, the interpolated error can be used as a correction function: the
values, which should be used in fitting of Helmholtz energy vs. volume curve,
would be $\{(V^{uc}, F^{uc}(V^{uc}) + \delta F(V^{uc}))_i\}$. Thus the effect
can be directly observed in numbers. The advantage of the former approach is
that the effect can be tackled analytically by means of calculus.

\section{Discussion}
\label{sec:discussion}
Although the theoretical considerations were presented for the Helmholtz energy,
completely analogous ideas are applicable also to some other thermodynamic
properties (e.g. internal energy, entropy), as the terms affected by the phonon
dispersion enter the respective expressions in a similar mathematical form
\cite{Anderson1995}. Certainly the negligible sensitivity of these quantities to
the phonon dispersion is connected with the peculiarities of their mathematical
definitions.

Another important aspect is the overwhelming role of static lattice energy (due
to its magnitude) in such properties as internal energy, bulk modulus and
others. As the consequence, the computational models are robust against
uncertainties in the description of lattice dynamics up to a good precision,
which is characterized by the relative error in calculations. Other properties,
e.g. the thermal expansion of solids, which is very small quantity compared to
compressibility, require more accurate techniques. Therefore its determination
is quite challenging task (not only for computational methods, but also for
measurements).

The effect of phonon dispersion may be rather more important to the quantities,
which require theoretical models of high accuracy, like thermal expansion or
Gruneisen parameter. Perhaps, one of the reason is that they are more closely
related to the microscopic properties and description of the system.
Besides, both thermal expansion and Gruneisen parameter are well known to be
affected by the anharmonicity of vibrations (moreover the anharmonicity plays a
crucial role in the phenomenon of thermal expansivity). In its turn, the
anharmonicity is an important factor that determines the phonon density of
states along with the phonon dispersion. And thus it raises altogether a
question on the interaction of both concepts and the level of approximation
achieved when one of the two is neglected.

In our opinion the materials exposed in this manuscript provide with convincing
arguments which lead to conclusions: i) the inclusion of phonon dispersion
into the computational model allows for systematic improvement of accuracy; ii)
the negligence of phonon dispersion does not deprives the model of the validity
in presence of other theoretical assumptions, moreover the loss of accuracy can
be expected to be negligible for many problems of concern.

\newpage

\begin{table}
\label{tbl:1}
\caption{Phonon dispersion integral of acoustic modes obeying Kieffer's model
for various parameters and temperatures.}
\begin{tabular}{c | c c}
\hline
\multirow{2}{*}{Parameter $b_{ac}$, Hz} &
\multicolumn{2}{c}{Phonon dispersion integral, $\mathrm{Ha} \cdot (3 p)^{-1}$}
\\
 & 300 K & 3000 K \\
\hline \hline
$10^{15}$         &  0.0671 &  0.0671 \\
$10^{13}$         &  0.0004 & -0.0188 \\
$5\cdot{}10^{12}$ & -0.0003 & -0.0254 \\
$3\cdot{}10^{12}$ & -0.0008 & -0.0302 \\
$10^{12}$         & -0.0019 & -0.0406 \\
$10^{11}$         & -0.0041 & -0.0625 \\
$10^{10}$         & -0.0062 & -0.0844 \\
$10^{9}$          & -0.0084 & -0.1063 \\
$1$               & -0.0281 & -0.3032 \\
\hline
\end{tabular}
\end{table}


\begin{thebibliography}{00}
\bibitem{Maradudin1963}
  A.A. Maradudin, E.W. Montroll and G.H. Weiss, Theory of Lattice Dynamics in
  the Harmonic Approximation; Academic Press: New York, London, 1963.
\bibitem{Evarestov2007}
  R.A. Evarestov, Quantum Chemistry of Solids: The LCAO first principles
  Treatment of Crystals; Springer: Berlin, Heidelber, New York, 2007.
\bibitem{Cotton1971}
  F.A. Cotton, Chemical Applications of Group Theory; Wiley: 3rd
  edition, 1990.
\bibitem{Anderson1995}
  O.L. Anderson, Equations of State of Solids for Geophysics and Ceramic
  Science; Oxford University Press: New York, Oxford, 1995.
\bibitem{Kieffer1979}
  S.W. Kieffer, {\it Rev. Geophys. and Space Phys.} 1979, 17, 1
\end{thebibliography}
\end{document}